\documentclass[a4paper,11pt]{article}
\usepackage{aaskaiid}
\usepackage{physics}
\usepackage{CJKutf8}  % add after acceptance
\usepackage{orcidlink}

\begin{document}

    \newcommand{\actaa}{Acta Astron.} % Acta Astronomica
    \newcommand{\araa}{ARA\&A} % Annual Review of Astron and Astrophys
    \newcommand{\aar}{A\&ARv} % Astrononmy \& Astrophysics Review
    \newcommand{\aapr}{A\&ARv} % Astronomy\&Astrophysics Reviews
    \newcommand{\ab}{Astrobiol.} % Astrobiology
    \newcommand{\aj}{AJ} % Astronomical Journal
    \newcommand{\apj}{ApJ} % Astrophysical Journal
    \newcommand{\apjl}{ApJL} % Astrophysical Journal, Letters
    \newcommand{\apjs}{ApJSS} % Astrophysical Journal, Supplement
    \newcommand{\ao}{Appl. Opt.} % Applied Optics
    \newcommand{\apss}{Astro. \& Space Sci.} % Astrophysics and Space Science
    \newcommand{\aap}{A\&A} % Astronomy and Astrophysics
    \newcommand{\aaps}{A\&AS.} % Astronomy and Astrophysics, Supplement
    \newcommand{\baas}{Bull. Am. Astron. Soc.} % Bulletin of the AAS
    \newcommand{\caa}{Chinese A\&A} % Chinese Astronomy and Astrophysics
    \newcommand{\cjaa}{Chinese J. A\&A} % Chinese Journal of Astronomy and Astrophysics
    \newcommand{\cqg}{Class. Quantum Gravity} % Classical and Quantum Gravity
    \newcommand{\gal}{Galaxies} % Galaxies
    \newcommand{\gca}{Geo. Cosmo. Acta} % Geochimica Cosmochimica Acta
    \newcommand{\icarus}{Icarus} % Icarus
    \newcommand{\jcap}{JCAP} % Journal of Cosmology and Astroparticle Physics
    \newcommand{\jgr}{J. Geophys. Res.} % Journal of Geophysics Research
    \newcommand{\jgrp}{J. Geophys. Res. Planets} % Journal of Geophysics Research: Planets
    \newcommand{\jqsrt}{J. Quant. Spectrosc. Radiat. Transf.} % Journal of Quantitiative Spectroscopy and Radiative Transfer
    \newcommand{\memsai}{Mem. SAIt} % Mem. Societa Astronomica Italiana
    \newcommand{\mnras}{MNRAS} % Monthly Notices of the RAS
    \newcommand{\nat}{Nature} % Nature
    \newcommand{\nastro}{Nat. Astron.} % Nature Astronomy
    \newcommand{\ncomms}{Nat. Commun.} % Nature Communications
    \newcommand{\nphys}{Nat. Phys.} % Nature Physics
    \newcommand{\na}{New Astron.} % New Astronomy
    \newcommand{\nar}{New Astron. Rev.} % New Astronomy Review
    \newcommand{\physrep}{Phys. Rep.} % Physics Reports
    \newcommand{\pra}{Phys. Rev. A} % Physical Review A: General Physics
    \newcommand{\prb}{Phys. Rev. B} % Physical Review B: Solid State
    \newcommand{\prc}{Phys. Rev. C} % Physical Review C
    \newcommand{\prd}{Phys. Rev. D} % Physical Review D
    \newcommand{\pre}{Phys. Rev. E} % Physical Review E
    \newcommand{\prx}{Phys. Rev. X} % Physical Review X
    \newcommand{\prl}{Phys. Rev. Let.} % Physical Review Letters
    \newcommand{\psj}{Planet. Sci. J.} % Planetary Science Journal
    \newcommand{\planss}{Planet. Space Sci.} % Planetary Space Science
    \newcommand{\pnas}{Proc. Natl Acad. Sci. USA} % Proceedings of the US National Academy of Sciences
    \newcommand{\procspie}{Proc. SPIE} % Proceedings of the SPIE
    \newcommand{\pasa}{PASA} % Publications of the Astron.  Soc. of Australia
    \newcommand{\pasj}{PASJ} % Publications of the Astron.  Soc. of Japan 
    \newcommand{\pasp}{PASP} % Publications of the Astron.  Soc. of the Pacific
    \newcommand{\rmxaa}{RMXAA} % Revista Mexicana de Astronomia y Astrofisica
    \newcommand{\sci}{Science} % Science
    \newcommand{\sciadv}{Sci. Adv.} % Science Advances
    \newcommand{\solphys}{Sol. Phys.} % Solar Physics
    \newcommand{\sovast}{Soviet Ast.} % Soviet Astronomy
    \newcommand{\ssr}{Space Sci. Rev.} % Space Science Reviews
    \newcommand{\uni}{Universe} % Universe

\setlength{\bibsep}{0.0pt} % separation between bib items

%\usepackage[backend=biber,style=authoryear]{biblatex}
%\addbibresource{refs.bib}

%\DeclareNameAlias{default}{family-given} 

\setcounter{secnumdepth}{5}
\makeatletter
\newcommand{\subsubsubsection}{\@startsection{paragraph}{4}{\z@}%
{1.5\baselineskip \@plus.5\dp0 \@minus.2\dp0}%
{.5\baselineskip \@plus2.3\dp0}%
{\reset@font\normalsize\itshape}
}
\newcommand{\subsubsubsubsection}{\@startsection{subparagraph}{5}{\z@}%
{1.5\baselineskip \@plus.5\dp0 \@minus.2\dp0}%
{.5\baselineskip \@plus2.3\dp0}%
{\reset@font\normalsize\itshape}
}
\makeatother
\setcounter{tocdepth}{5}

\title{Search for Seeds of The Cradle of Stars: Study of The Evolution From Atomic to Molecular Hydrogen Clouds with 1000 AU Scale Resolution }
\ShortTitle{CNM Formation at 1000 AU Scale}

\author[1]{Hiroaki Yamamoto \begin{CJK*}{UTF8}{ipxm}(山本宏昭)\end{CJK*} \orcidlink{0000-0001-5792-3074}}
\ShortName{Yamamoto et al.} % shortened name list for header 
\author[1]{Yamato Matsuzuki \begin{CJK*}{UTF8}{gbsn}(松月大和)\end{CJK*}}
\author[1]{Kengo Tachihara  \begin{CJK*}{UTF8}{gbsn}(立原研悟)\end{CJK*} \orcidlink{0000-0002-1411-5410}}

\affiliation[1]{Department of Physics, School of Science, Nagoya University, Furo-cho, Chikusa-ku, Nagoya, Aichi, Japan, 464-8602}
\emailAdd{hiro@a.phys.nagoya-u.ac.jp}

\abstract{The aim of this science case for SKA1-Mid is to elucidate the spatial and velocity structure of the Cold Neutral Medium (CNM) on scales below the Field length, thereby elucidating the physical processes that govern CNM structure formation and its role in the earliest stages of molecular cloud formation. 
This goal will be achieved through a unique observational combination of several-arcsecond angular resolution, high velocity resolution, and unprecedented surface-brightness sensitivity. 
These observations will reveal the characteristic size, morphology, velocity structure, and internal dynamics of CNM clumps, and will directly connect such structures to the earliest phases of the atomic to molecular transition. 
Full-polarization observations with SKA1-Mid will also enable emission-based Zeeman measurements of the CNM, allowing the magnetic field strength to be mapped and opening a new window on the role of magnetic fields in CNM formation and evolution. 
In particular, this will make it possible to determine whether magnetic fields primarily inhibit fragmentation by supporting the gas against compression, or instead guide anisotropic condensation and promote the development of small-scale CNM structure through thermal instability. 
In combination with interferometric spatial filtering and high spectral resolution, SKA1-Mid will selectively trace the CNM and robustly identify individual clumps in both position and velocity space. 
These capabilities will provide decisive tests of thermal instability and the two-phase turbulence model, and will establish a new observational foundation for understanding how molecular clouds emerge from the multi-phase atomic interstellar medium.}

%% \tableofcontents
\maketitle

\section{Introduction}
Hydrogen constitutes the principal component of the interstellar medium (ISM), existing in ionized, neutral atomic, and molecular forms. 
Neutral atomic hydrogen (H\,\textsc{i}) is typically classified into three distinct components: the warm neutral medium (WNM; high temperature and low density), the cold neutral medium (CNM; low temperature and high density), and the lukewarm neutral medium (LNM), a thermally unstable state that lies between the two \citep{McClure-Griffiths2023}. 
The CNM is considered to form through thermal instability acting on the WNM, triggered by shocks and turbulence, and represents an intermediate stage in the transition from diffuse atomic cloud to molecular cloud.
Observationally, analyses of all-sky H\,\textsc{i} clouds have estimated the mean fractions of the CNM, LNM, and WNM in the local H\,\textsc{i} clouds to be $\sim$25\%, $\sim$41\%, and $\sim$34\%, respectively. 
These fractions in the H\,\textsc{i} filaments are $\sim$46\% for the CNM, $\sim$37\% for the LNM, and $\sim$17\% for the WNM, although the CNM fraction is a lower limit because no correction for optical depth has been applied. 
These results indicate that the relative contributions of the three phases vary depending on the region and/or the evolutionary stage \citep{Kalberla2018}.

This transition is of fundamental importance because molecular cloud provides the immediate reservoir for star formation \citep[e.g.,][]{Krumholz2009a, Kennicutt2012}. 
Although molecular hydrogen can form under a range of interstellar conditions, efficient H$_2$ formation generally requires atomic hydrogen cloud to become sufficiently cold, dense, and shielded from dissociating ultraviolet radiation \citep[e.g.,][]{Krumholz2009b}. 
The CNM therefore plays a key role as a primary pathway by which diffuse atomic cloud is compressed and cooled before becoming molecular \citep{Koyama2002, Wolfire2003}. 
CNM formation may not be necessarily the only possible pathway to molecular cloud, especially in dynamically compressed or non-equilibrium environments. 
However, in the local Galactic ISM, molecular cloud formation generally requires the production of cold, dense, and shielded atomic gas, so the CNM or CNM-like thermally unstable gas is expected to be a major, and probably primary, intermediate stage \citep{Wolfire2003, Inoue2012, Kobayashi2020}. 
In this picture, the WNM first condenses into thermally unstable LNM, and then into dense CNM structures, which can subsequently accumulate sufficient column density and shielding to enable the formation of H$_2$ and, eventually, molecular clouds \citep{Koyama2002, Inoue2012, Kobayashi2020}. 
Understanding the spatial structure, kinematics, and physical conditions of the CNM is therefore essential for clarifying how the atomic ISM evolves into the molecular phase.

Absorption line studies toward compact background continuum sources have provided valuable constraints on the physical properties of the CNM  \citep[e.g.,][]{Heiles2003a, Stanimirovic2014, Murray2015}.  
However, because such measuremets are limited to discrete line of sight such as quasars, they are unable to elucidate extended spatial structure of the CNM clouds. 
Temporal variations in the optical depth of absorption line caused by the proper motion or annual parallax of background sources have revealed the  structures of the H\,\textsc{i} clouds on scales of 10--100 AU \citep[e.g.,][]{Faison2001, Stanimirovic2018}. 
While these results demonstrate that the CNM exhibits structure on very small scales, the spatial coverage remains confined to the narrow regions sampled by background sources, preventing a comprehensive view of CNM morphology and kinematics.

A key physical scale in the CNM is the Field length associated with thermal instability, which is an order of $\sim$0.01 pc ($\sim$2800 AU).
This implies that observations capable of resolving spatial scales down to 1000 AU are crucial for testing theoretical models of CNM formation. 
To capture the spatial distribution of the H\,\textsc{i} clouds, emission line observations are indispensable. 

The increasing sensitivity of modern radio interferometers will substantially raise the number of detectable background sources. 
In the MeerKAT MIGHTEE survey, $\sim$114,000 sources have been detected over an area of $\sim$20.1 deg$^{2}$ with an angular resolution of 5$^{\prime\prime}$, and a sensitivity of 1--3.6 $\mu$Jy beam$^{-1}$ \citep{Hale2025}. 
This corresponds to an average source number density of $\sim$4.4$\times$10$^{-4}$ arcsec$^{-2}$. 
Furthermore, in the SKA era, the number of background point sources detectable at 1 GHz is expected to reach $\sim$65,000 deg$^{-2}$ at a sensitivity of 1 $\mu$Jy, equivalent to $\sim$5$\times$10$^{-3}$ arcsec$^{-2}$ \citep{Jarvis2015}. 
This corresponds to only $\sim$0.24 background sources per 1000 AU $\times$ 1000 AU area at a distance of 150 pc, indicating that, although the situation is improved compared to MeerKAT, the sampling remains intrinsically sparse on such scales.
Consequently, absorption line observations alone will be insufficient to fully sample the spatial distribution of the CNM, and emission line observations are essential for mapping its extended structure.

Numerical simulations of molecular cloud formation predict that the CNM emerges as a population of compact clumps with multiple velocity components that overlap both spatially and spectrally \citep[e.g.,][]{Inoue2012, Kobayashi2020}. 
These results imply that high angular resolution, high spectral resolution, and high surface brightness sensitivity are all required to disentangle CNM substructure, and to connect observed properties with theoretical models. 
SKA1-Mid will uniquely provide these capabilities, enabling systematic studies of CNM structure and kinematics down to $\sim$1000 AU scales over spatially extended regions.

Interstellar magnetic fields exert a profound influence on the structure and evolution of the ISM. 
Numerical simulations show that variations in magnetic field strength lead to markedly different evolutionary pathways, underscoring the dominant role of magnetic fields in shaping interstellar structure and dynamics \citep[e.g.,][]{Kortgen2015}.

The Zeeman effect \citep{Zeeman1897} provides a unique tool for measuring interstellar magnetic fields. By fitting the Stokes V spectrum produced by the Zeeman effect, the strength of the interstellar magnetic field can be directly determined \citep{Crutcher2019, Pattle2023}. 
The Zeeman splitting coefficient for H\,\textsc{i} is 2.8 Hz $\mu$G$^{-1}$. 
Accounting for Doppler shifts, H\,\textsc{i} exhibits a comparatively large frequency displacement per unit velocity, making it a particularly effective probe of magnetic field strengths. 
However, overlapping emission components along the line of sight have historically limited Zeeman measurements in emission, leading most studies to rely on absorption line observations \citep[e.g.,][]{Heiles2005}. 
Nevertheless, recent observations have demonstrated that Zeeman measurements using H\,\textsc{i} emission can be achieved in carefully selected regions. \citet{Ching2022} measured the line-of-sight magnetic field toward the L1544 region using H\,\textsc{i} emission and H\,\textsc{i} narrow self-absorption (HINSA) data. 
This demonstrates that emission-based H\,\textsc{i} Zeeman measurements are feasible when the relevant cold components can be sufficiently isolated. 
In addition, absorption measurements cannot deliver spatially resolved magnetic field maps within the CNM. 
Therefore, in order to map the spatial distribution of magnetic fields across the CNM, it is essential to establish Zeeman measurements in H\,\textsc{i} emission as a practical observational technique. 
In this work, we outline a science case in which advanced spectral decomposition techniques are used to isolate individual CNM components in emission, and we demonstrate, using existing data and simulations, that such an approach can enable robust Zeeman measurements in the SKA era. 
This capability will provide a fundamentally new observational window on the interplay between thermal instability, turbulence, and magnetic fields in shaping the cold atomic phase of the ISM.

\section{Science Goal at SKA1-Mid}

\subsection{How Does Thermal Instability Promote the Formation of the CNM ?}

Thermal instability is widely considered to be one of the key mechanisms responsible for the emergence of the cold neutral medium (CNM) from diffuse atomic gas.
In the classical picture, small perturbations in thermally unstable H\,\textsc{i} gas can grow through radiative cooling, leading to the condensation of cold structures embedded within a warmer ambient medium.
The minimum scale on which such condensations can survive against thermal conduction is given by the Field length \citep{Field1965}, $\lambda$,
\begin{equation}
\lambda = \sqrt{\frac{KT}{n^2\Lambda}},
\end{equation}
where $K$ denotes the thermal conductivity of hydrogen atoms, $\Lambda$ is the cooling function, and $n$ and $T$ are the number density and temperature of the H\,\textsc{i} gas, respectively.
Adopting $K = 2.5 \times 10^3 T^{0.5}$ erg cm$^{-1}$ K$^{-1}$ s$^{-1}$ \citep{Parker1953} and $\Lambda = 5.7 \times 10^{-26} (T/10^4)^{0.8}$ erg cm$^3$ s$^{-1}$ \citep{Wolfire2003}, the Field length is estimated to be $\sim$2800 AU for typical CNM conditions of $T = 100$ K and $n = 30$ cm$^{-3}$.
This scale is of particular interest because it links the microphysics of cooling and conduction to the observable structure of the atomic interstellar medium.
More fundamentally, it marks the characteristic boundary scale at which thermal conduction and radiative cooling balance each other, so that the growth of thermal instability is suppressed below it.
The Field length is therefore not merely a small spatial scale within a turbulent hierarchy, but a physically motivated condensation scale that sets the minimum size of long-lived CNM structures.

More recent theoretical work has extended this classical picture to a turbulent, multiphase medium.
The two-phase turbulent model provides a framework in which the small-scale structure of the atomic interstellar medium emerges from the coupled effects of turbulence and thermally driven phase transition.
In this scenario, turbulent compression within the WNM triggers transitions into the thermally unstable regime, followed by rapid cooling and condensation into CNM.
Because these phase transitions occur in a dynamical environment, the CNM is expected to exhibit complex spatial and velocity structures, including sharp density contrasts, fragmented morphology, and non-uniform internal motions.
Testing these predictions requires observations that can simultaneously resolve the spatial and kinematic structure of the CNM \citep[e.g.,][]{Koyama2002,Hennebelle2007,Inoue2012,Saury2014}.
Within this framework, the Field length represents the minimum characteristic scale at which thermally unstable perturbations can condense into long-lived CNM structures, while turbulent motions continuously reshape, merge, and disperse them.
The resulting CNM is therefore expected to retain signatures of both thermal instability and turbulence, making the observational characterization of its characteristic scales an important test of multiphase ISM theory.

This scale regime is also directly relevant to the broader problem of molecular cloud formation.
If CNM structures formed near the Field length already contain significant internal motions, large density contrasts, or persistent anisotropic morphologies, then they may provide the seeds from which denser molecular structures subsequently develop.
In that case, the physical state of the CNM at its formation scale would set the initial conditions for later stages of molecule formation and gravitational fragmentation.
Conversely, if the smallest CNM condensations are controlled primarily by thermal physics and remain dynamically simple, then turbulence may become dominant only at somewhat larger scales.
Determining the characteristic size, morphology, and internal kinematics of CNM condensations is therefore essential for understanding not only how thermal instability operates, but also how the atomic medium evolves toward the conditions required for molecular cloud formation.

Observationally, however, this regime has remained poorly explored.
The characteristic scale associated with the Field length is far below that resolved in most existing H\,\textsc{i} studies, and the physical properties of CNM structures at these scales are still uncertain.
As described in Section 3.1, we adopt a synthesized beam of 4$^{\prime\prime}$.84 $\times$ 3$^{\prime\prime}$.59, which corresponds to $\sim$700 AU $\times$ $\sim$520 AU at a source distance of 150 pc. This resolution is sufficient to resolve the Field length in nearby atomic structures. A distance of 150 pc is adopted as a representative value because most of the target clouds are located at approximately this distance.
Further details are given in Section 3.3.
At this resolution, the surface distribution of the CNM can be characterized quantitatively in terms of its column density, morphology, and large-scale kinematics.
For individual CNM structures, it also becomes possible to probe internal linewidths down to $\sim$0.2 km s$^{-1}$ and to infer associated physical conditions such as densities ($n$ $\sim$10$^{1-2}$ cm$^{-3}$) and temperatures ($T$ $\sim$100 K).
These measurements will provide direct constraints on the scales and internal properties of CNM clumps that have so far been discussed primarily on theoretical grounds.

This point is closely related to the question of whether the CNM near its condensation scale already exhibits the dynamical signatures that later characterize molecular clouds.
Observations of low-density molecular clouds traced by the $^{12}$CO($J$=1--0) line almost always show linewidths larger than expected from thermal broadening alone, indicating that even young molecular structures are already turbulent.
This suggests that turbulence is already present in the CNM before the onset of molecule formation \citep{Matsuzuki2026}.
For an H\,\textsc{i} cloud at $T$ $\sim$100 K, the thermal linewidth is $\sim$2.1 km s$^{-1}$, corresponding to a velocity dispersion of $\sim$0.9 km s$^{-1}$.
Because SKA1-Mid can resolve both the Field length and the thermal linewidth of the CNM, it provides an opportunity to investigate how thermal instability and turbulent motions combine to produce the cold atomic structures that may later evolve into molecular clouds.
These observations will also enable a direct test of the two-phase turbulence models, in which the CNM and WNM are dynamically coupled but exhibit distinct spatial and kinematic characteristics arising from thermal instability and shock compression.
By resolving CNM structures on sub-parsec and sub-Field length scales, SKA1-Mid will make it possible to examine whether the cold atomic gas already preserves the characteristic signatures of phase-coupled turbulence before molecule formation begins.
Such observations will help clarify whether the CNM on these scales already contains the dynamical seeds of molecular cloud formation, how those seeds emerge from the interplay between thermal instability and turbulence, and whether they are consistent with the predictions of the two-phase turbulent ISM framework.

\subsection{What Role Does the Interstellar Magnetic Field Play in the Formation of the CNM?}

Magnetic fields are a pervasive component of the Galactic interstellar medium and are widely expected to influence the formation and evolution of cold atomic gas.
All-sky dust polarization maps reveal ordered magnetic field structures over a broad range of Galactic environments \citep{PlanckCollaboration2018}, while H\,\textsc{i}4PI data demonstrate that atomic hydrogen is distributed throughout the sky \citep{BenBekhti2016}.
In addition, \citet{Booth2026} used Faraday depth spectra over 500--1030 MHz to reveal the three-dimensional distribution of the local large-scale interstellar magnetic field, showing that much of the observed Faraday sky may be dominated by magnetic structure within $\sim$1 kpc of the Sun.
These observations imply that magnetic fields and atomic gas coexist essentially everywhere in the Galaxy.
Nevertheless, the quantitative relationship between magnetic fields and the physical properties of the CNM remains poorly constrained.

From a theoretical perspective, magnetic fields can influence CNM formation in several ways.
They can modify the compressibility of shocked gas, channel converging flows, alter the growth rate of thermal instability, and affect the morphology and survival of condensed structures \citep[e.g.,][]{Hennebelle2019}.
Numerical studies suggest that the efficiency of CNM condensation, the characteristic sizes of CNM structures, and their subsequent transition toward molecular clouds depend sensitively on both the strength and orientation of the magnetic field \citep[e.g.,][]{Heitsch2009}.
More specifically, magnetohydrodynamic simulations indicate that magnetic fields can strongly inhibit or channel compression depending on the geometry of the converging flow relative to the mean field direction.
In particular, magnetic pressure can hinder strong compression when converging motions are largely perpendicular to the field, while permitting condensation to proceed more effectively under more favorable geometrical conditions \citep{Inoue2008,Inoue2009}.
Magnetic fields therefore regulate not only the efficiency of condensation, but also the anisotropy and morphology of the structures that subsequently develop.

More recent simulations have reinforced and refined this picture.
In colliding magnetized flows, the magnetic field can regulate fragmentation by suppressing motions perpendicular to the field, thereby altering or even weakening the formation of large-scale filamentary substructures, although molecular clump and core formation can still proceed \citep{Weis2024}.
In addition, simulations of cloud formation in colliding flows show that the relative orientation between magnetic fields and density structures is not fixed, but can change systematically with density and magnetization, including transitions from predominantly parallel to predominantly perpendicular configurations at high density \citep{Seifried2020}.
Complementary work on thermally bistable MHD flows further suggests that the alignment between CNM clouds and magnetic fields is closely linked to the dynamics of shock compression itself: fast MHD shocks can generate field fluctuations, which are then amplified by downstream compressive velocity gradients, providing a physical mechanism for the observed relative orientations of magnetic fields and CNM structures \citep{GrandaMunoz2025}.
Taken together, these results indicate that magnetic fields do not merely accompany CNM formation, but actively shape its fragmentation, morphology, and dynamical evolution.

These theoretical results imply that magnetic fields are likely to influence the earliest stages of molecular cloud formation as well.
Because the CNM is the immediate precursor of denser molecular gas, any magnetic regulation of CNM condensation, fragmentation, or alignment may leave an imprint on the structure and evolution of the gas that later becomes molecular.
The geometry of the field relative to converging flows may determine not only whether cold structures form efficiently, but also whether they evolve into sheets, filaments, or more fragmented condensations.
Magnetic regulation at the CNM stage may therefore help set the initial conditions for the subsequent development of molecular clouds, making observational tests of magnetic field strength and morphology in the atomic phase especially important.

Observational constraints on magnetic fields in atomic gas have, however, been limited.
Most direct measurements rely on Zeeman splitting, either in absorption toward bright background sources \citep{Heiles2004} or, more indirectly, through comparisons with dust-based tracers \citep{PlanckCollaboration2016}.
While these studies have provided valuable insight, they remain sparse and do not easily yield spatially extended measurements of magnetic field strength in the cold atomic phase itself.
In particular, dust polarization probes somewhat different density and temperature regimes and depends on assumptions about grain alignment and gas--dust coupling.
Absorption-line Zeeman observations provide more direct information, but are restricted to isolated sightlines and therefore offer limited spatial coverage.
According to the review by \citet{Heiles2005}, the characteristic magnetic field strength in diffuse clouds is $6.0 \pm1.8 \,\mu{\rm G}$, underscoring both the astrophysical importance of the field and the observational difficulty of measuring such weak signals directly.

Recent LOFAR studies have suggested that Faraday tomography can probe magnetic fields associated with the multiphase neutral ISM. 
\citet{Bracco2020} found significant morphological correlations between LOFAR Faraday-depth structures and EBHIS H\,\textsc{i} emission in high-Galactic-latitude fields. 
Using ROHSA to decompose the H\,\textsc{i} spectra into CNM, LNM, and WNM components, they showed that the correlation is particularly strong for the CNM and LNM, indicating a close connection between the magneto-ionic medium and phase transition in diffuse H\,\textsc{i}. 
However, because the ionization fraction and path length of the CNM are generally too small to explain the observed Faraday depths by the CNM alone, this correlation should be interpreted as an indirect tracer of the magnetized environment surrounding CNM structures rather than as a direct measurement of CNM magnetic fields.

This interpretation is supported by \citet{Boulanger2024}, who combined LOFAR Faraday spectra, Planck dust polarization, and FUSE ultraviolet absorption-line measurements and showed that a slab of magnetized, partially ionized WNM in the local ISM can account for the observed LOFAR Faraday structures. 
Furthermore, \citet{Berat2026} used MHD simulations of thermally bistable H\,\textsc{i} gas and synthetic LOFAR observations to show that thermal electrons associated with neutral H\,\textsc{i} can reproduce Faraday depths and polarized intensities comparable to those observed in the 3C196 field. 
Their results also indicate that CNM structures can be morphologically correlated with Faraday tomographic features, although the relative contributions of CNM and WNM depend on turbulence, magnetic-field orientation, observational noise, and line-of-sight geometry.

These studies suggest that Faraday tomography is a promising indirect probe of magnetic-field morphology in the multiphase neutral ISM. 
The WNM is likely to provide a substantial fraction of the Faraday-rotating material because of its larger ionization fraction, whereas the CNM traces the cold, dense structures that are dynamically coupled to this magnetized WNM/LNM environment. 
Thus, Faraday tomography can link the magnetic field of partially ionized WNM to the formation and morphology of CNM filaments, but it does not by itself yield a direct measurement of the magnetic-field strength inside the CNM. 
It is therefore complementary to H\,\textsc{i} Zeeman observations, which can directly measure the line-of-sight magnetic field in individual H\,\textsc{i} components.

The use of H\,\textsc{i} emission to trace Zeeman splitting has long been recognized as potentially powerful, but observationally challenging.
Emission spectra often contain multiple velocity components blended along the line of sight, which dilute the Zeeman signature of individual CNM structures.
Moreover, such measurements require both high sensitivity and very accurate polarization calibration (see Section 3.2).
Even so, previous work has demonstrated that emission-based measurements can be carried out in favorable cases.
\citet{Heiles1998,Heiles1999} reported magnetic field measurements from the Zeeman effect in H\,\textsc{i} emission toward high Galactic latitude molecular clouds, showing that the method is feasible when the target morphology and instrumental systematics are carefully controlled.
For this reason, emission-based magnetic field measurements in the CNM should be regarded as difficult but achievable with sufficiently sensitive and well-calibrated facilities.
We have recently demonstrated that H\,\textsc{i} line components can be successfully separated through Gaussian decomposition, and we adopt this technique here as a spectral decomposition method \citep{Matsuzuki2026}.
The sensitivity and full-polarization capability of SKA1-Mid, combined with this spectral decomposition approach to isolate individual CNM components in both velocity and space, offer a new route forward.
By mitigating the effects of line blending, this method can make it possible to recover weak Zeeman signatures directly from H\,\textsc{i} emission.

A particularly important aspect of this approach is that it enables spatially extended measurements rather than isolated line of sight estimates.
This makes it possible to investigate how magnetic field strength varies across the CNM and how it correlates with density, temperature, velocity dispersion, and spatial scale.
Such information is needed to assess whether magnetic fields primarily suppress, channel, or enhance CNM condensation driven by thermal instability, and to compare the relative importance of magnetic, thermal, and turbulent pressures in the cold atomic medium.
Because H\,\textsc{i} emission is available over the entire sky, this methodology may ultimately provide a basis for comparative studies of CNM magnetization across a wide variety of Galactic environments, from diffuse high-latitude clouds to the atomic envelopes of molecular clouds.
In this sense, the combination of SKA1-Mid sensitivity, angular resolution, and polarization fidelity may open a new observational window on the role of magnetic fields in structuring the cold atomic ISM and setting the initial conditions for molecular cloud formation.

\subsection{Why Is Spatially Resolved H\,\textsc{i} Imaging Essential for Understanding the Origin of CNM Structure?}

A major obstacle in understanding the origin of small-scale CNM structure is that existing studies have relied predominantly on H\,\textsc{i} absorption measurements toward compact background sources \citep[e.g.,][]{Heiles2003a}.
These observations have been crucial in establishing the existence of tiny-scale opacity fluctuations in atomic gas \citep{Stanimirovic2018}, but they provide only pencil-beam information along isolated sightlines.
As a result, they cannot recover the two-dimensional morphology of the gas on the plane of the sky, nor can they determine whether the detected fluctuations correspond to compact cloudlets, elongated filaments, sheets, or line-of-sight superpositions.
This limitation is particularly severe when addressing the physical origin of CNM structure on scales approaching the Field length, where the first condensations of the cold atomic medium are expected to emerge.

This scale regime is of central importance for CNM physics.
Thermal instability provides the basic route by which a thermally bistable atomic medium can condense into a cold phase \citep{Field1965}, but CNM formation is not determined by thermodynamics alone: converging flows can nonlinearly trigger thermal instability and promote condensation \citep{Koyama2002,Inoue2012}, while the resulting fragmentation can generate turbulent motions and shape the spatial complexity of the cold gas \citep{Audit2005}.
Magnetic fields add a further level of control, because they can strongly inhibit or channel compression depending on the geometry of the converging flow relative to the mean field direction, thereby regulating the efficiency of condensation as well as the anisotropy and morphology of the resulting structures \citep{Inoue2008,Inoue2009}.
The key unresolved question is therefore not whether these processes matter, but which of them dominates the initial formation scale of the CNM and how their relative importance changes with spatial scale.

The Field length is not simply another point along a turbulent hierarchy, but the physically defined boundary scale below which thermal conduction suppresses the growth of thermal instability.
Resolving CNM structures on scales comparable to and below the Field length therefore gives direct observational access to the condensation scale itself.
Such observations can test whether the smallest CNM structures are fundamental products of thermal instability or instead reflect velocity and density structure inherited from larger-scale turbulent or magnetically guided flows.
Equally importantly, they can reveal the scale at which the velocity field becomes genuinely turbulence-dominated rather than being governed primarily by thermal instability, phase transition, or condensation-driven motions.

Spatially resolved H\,\textsc{i} imaging is therefore indispensable.
As described in Section 3.1, we adopt a synthesized beam of 4$^{\prime\prime}$.84 $\times$ 3$^{\prime\prime}$.59 ($\sim$700 AU $\times$ $\sim$520 AU), which reaches the scale required to resolve the Field length for nearby atomic structures.
By mapping the H\,\textsc{i} emission over an extended field with both high angular and high spectral resolution, one can isolate the CNM in both position and velocity and directly measure its projected size, aspect ratio, column density, and linewidth.
This provides a qualitatively distinct form of information from that obtainable through absorption-line studies alone.
In particular, such observations will make it possible to detect a large population of CNM clumps \citep[e.g.,][]{Matsuzuki2026}, thereby enabling a statistical characterization of CNM structures spanning a wide range of environments, spatial scales, and evolutionary stages within a single observational framework.
Instead of relying on a sparse collection of one-dimensional sightlines, one obtains a resolved population of CNM structures that can be studied in a unified way across the sky.

A particularly important diagnostic in this context is the linewidth--size relation \citep{Larson1981}.
For H\,\textsc{i} gas at $T$ $\sim$100 K, the thermal linewidth is already $\sim$2.1 km s$^{-1}$, whereas the non-thermal velocity dispersion expected from a simple extrapolation of Larson's relation to $\sim$10$^3$ AU is only $\sim$0.1--0.2 km s$^{-1}$.
Sparse absorption measurements are therefore fundamentally inadequate for distinguishing whether the observed velocity widths at these scales arise from turbulence, thermal broadening, or condensation-related motions.
Spatially resolved emission-line mapping changes this situation completely.
By measuring linewidths for a large ensemble of CNM structures as a function of size and environment, SKA1-Mid will make it possible not simply to ask whether a Larson-like relation extends to very small scales, but to determine how the character of the velocity field changes as one approaches the condensation scale of the CNM.
In this way, the linewidth--size relation becomes an empirical diagnostic of the CNM velocity field, rather than a direct proof of a fully developed turbulent cascade.

This observational capability is also crucial for connecting CNM structure to the broader problem of molecular cloud formation.
The CNM is the immediate precursor of molecular clouds, and the physical state of CNM condensations sets the initial conditions from which denser molecular structures subsequently develop.
Determining how thermal instability, turbulence, and magnetic fields shape CNM structure at the Field length scale is therefore essential for understanding how cold atomic gas evolves toward the conditions required for H$_2$ formation and gravitational fragmentation.
In this sense, resolving the CNM at its characteristic condensation scale is not simply a technical advance in H\,\textsc{i} imaging, but a decisive step toward establishing the physical origin of cold atomic structure and the initial conditions of molecular cloud formation.

Taken together, this science case defines a uniquely powerful role for SKA1-Mid.
It will provide the first spatially resolved view of CNM structure at and below the Field length, the first statistical access to the morphology and kinematics of CNM condensations at their formation scale, and the first direct opportunity to test whether thermal instability, turbulence, or magnetic regulation is the dominant driver of small-scale CNM structure.
By resolving the CNM in both space and velocity, SKA1-Mid will open a new observational window on how cold atomic structures emerge from the diffuse ISM and how the atomic medium is prepared for subsequent molecular cloud formation.
While the CNM with narrow line width is a plausible candidate for the cold atomic cloud associated with molecular cloud formation, comparison with tracer of molecular clouds such as CO or OH is essential to strengthen its identification as the component physically linked to the site of molecular cloud formation.

\section{Requirements for SKA-Mid observations}

\subsection{Sensitivity}

The objectives of the observation are the following two: (i) to detect CNM structures on $\sim$1000 AU scales, and (ii) to obtain a two-dimensional map of the interstellar magnetic field strength. 
Although the sensitivities required for these two goals differ, both can be achieved within a single observing program. 
The relationship between \textsl{Stokes I} and \textsl{Stokes V} arising from the Zeeman effect can be expressed as
\begin{equation}
    \frac{(Stokes\ V)}{(Stokes\ I)} = \frac{Z \times B_{LOS}}{\Delta \nu_{\rm line}}
\end{equation}

This formula follows from formula (3) of \citet{Crutcher2019}.
\citet{Heiles2005} reported a magnetic field strength of 6.0$\pm$1.8 $\mu$G in the CNM, while \citet{Boulanger2024} estimated the magnetic field strength in the WNM to be a few $\mu$G. 
These results suggest that the magnetic field strength in the regions targeted in this study is also expected to be on the order of a few $\mu$G. 
We therefore adopt a field strength of a few $\mu$G in our sensitivity calculation.
For magnetic fields of a few $\mu$G, $\frac{Stokes V}{Stokes I}$ is on the order of 10$^{-4}$.

As representative values for a detectable CNM component, we adopt an H\,\textsc{i} spin temperature of $T_{\mathrm{s}}$ $\sim$100 K and an optical depth of $\tau$ $\sim$0.1, which are broadly consistent with Galactic CNM components observed in H\,\textsc{i} emission--absorption studies \citep[e.g.,][]{Heiles2003a,Heiles2003b,Murray2018}.
For such a component detectable with SKA1-Mid AA4, the expected brightness temperatures are $\sim$10 K for \textsl{Stokes I} spectrum, and $\sim$0.05 K for \textsl{Stokes V} spectrum.
Based on our experience with ROHSA analyses, the phase separation of the H\,\textsc{i} degrades when the velocity resolution is coarser than 0.5 km s$^{-1}$. 
To detect CNM on 1000 AU scales while maintaining $\mathrm{S/N}$ $\sim$20, we estimated the sensitivity using the parameters in Table 1 with the sensitivity calculator. 
The synthesized beam is 4$^{\prime\prime}$.84 $\times$ 3$^{\prime\prime}$.59 ($\sim$700 AU $\times$ $\sim$520 AU), yielding a per-channel (spectral) surface-brightness sensitivity of $\sim$1.18 K. 
Applying spectral averaging by a factor of four in the analysis gives an effective resolution of 1.26 kHz (266.0 m s$^{-1}$), at which the per-channel surface-brightness sensitivity improves to $\sim$0.48 K.
A similar high spatial resolution and sensitivity can also be achieved with SKA1-Mid AA$^{*}$.
However, the required integration time is a$\sim$25\% longer than that for SKA1-Mid AA4.
Although MeerKAT can achieve a comparable angular resolution, its sensitivity is far from sufficient. 
Therefore, the present study is expected to be feasible only with SKA1-Mid.

For \textsl{Stokes V}, adopting a UV taper of 15$^{\prime\prime}$.775 results in a synthesized beam of 44$^{\prime\prime}$.24 $\times$ 31$^{\prime\prime}$.18 ($\sim$6400 AU $\times$ $\sim$4500 AU) and a per-channel surface-brightness sensitivity of $\sim$0.015 K. 
Thus, \textsl{Stokes V} is detectable at $\sim$3$\sigma$. 
Although the \textsl{Stokes V} beam is roughly an order of magnitude larger than that for \textsl{Stokes I}, analysis at this angular scale remains sufficiently informative. 
If SKA2 is realized, its resolution and sensitivity will be dramatically improved owing to 10 times better sensitivity than SKA1 \citep{Robert2019}.
It enables the derivation of \textsl{Stokes I} and \textsl{Stokes V} with a spatial resolution of 1000 AU within a realistic observing time. 
This will make it possible to obtain two-dimensional maps of magnetic field strength on the Field length scale.

\begin{table}[h]
	\centering
	\caption{Input parameters to sensitivity calculator}
	\label{tab:example_table}
	\begin{tabular}{ll} % four columns, alignment for each
		\hline
		T$_{\rm sys \ 15-m}$ & 20 K \\
		Center Frequency & 1420 MHz \\
		Band Width &  3.125MHz  \\
		Spectral Resolution & 0.21kHz \\
		Image weighting & Briggs \\
		Robust & +1 \\
		Tapering & 0$^{\prime\prime}$.986 \\
		Integ. time & 100 hours \\
		\hline
	\end{tabular}
\end{table}

\subsection{The Required Accuracy of Polarization Calibration}

For linearly polarized feeds, \textsl{Stokes V} is formed from the cross-hand correlations rather than from the difference between the two parallel hands. 
Therefore, differential gain and phase errors between the two linear receptors, as well as polarization leakage terms, can bias the recovered \textsl{Stokes V}. 
In particular, residual cross-hand phase errors mix \textsl{Stokes U} and \textsl{V}, which is critical for Zeeman measurements.
For H\,\textsc{i} Zeeman measurements of magnetic field strength of a few $\mu$G, the expected fractional level of \textsl{Stokes V} is only $\frac{V}{I}$ $\sim$10$^{-4}$.
Instrumental leakage into \textsl{Stokes V} must therefore be controlled below this level. 
A relative gain stability below 0.1\% and a residual cross-hand phase error of 0.05$^\circ$ are desirable for robust measurements. 
For linearly polarized feeds, the cross-hand phase calibration is particularly important because residual phase errors mix \textsl{Stokes U} and \textsl{V}.
Achieving these requirements critically depends on the choice of calibrator, and the calibration cadence must be increased relative to that of standard observing procedures.

\subsection{Candidates of the target on this study}

To accomplish the objectives of this study, a spatial resolution better than 1000 AU is required. 
As discussed in Section 3.2, from the viewpoint of the sensitivity achievable within a realistic observing time, this requirement can only be met for targets located within a few hundred pc, for which an angular resolution of $\sim$4 arcsec is needed. 
In addition, low Galactic latitude regions suffer from severe line-of-sight contamination in H\,\textsc{i} emission, making high-Galactic-latitude targets, typically at $|b| \gtrsim$ several degrees, more suitable for this study. 

Nearby star-forming regions, including Taurus, Lupus, and Ophiuchus, are associated with the Local Bubble and are located at distances of $\sim$150 pc \citep{Zucker2022}. 
Distances have also been estimated for several high Galactic latitude molecular clouds. 
MBM 53, 54, and 55, and the Pegasus Loop are located at $\sim$150 pc and $\sim$100 pc, respectively \citep{Yamamoto2003, Yamamoto2006}. 
The Galactic coordinates of these objects are provided in the references cited above. 
These nearby objects provide suitable targets for the present study because they meet the distance requirement needed to resolve the Field length, and their high Galactic latitudes help minimize line of sight contamination in H\,\textsc{i} emission. 
Targeting such objects at distances of order 150 pc therefore enables us to spatially resolve the Field length and investigate the physical origin of CNM formation.

The objects listed above are primarily molecular cloud regions. 
In particular, for objects in the solar neighborhood, stars provide the primary distance indicators, and distance to molecular clouds have historically been constrained by those stars associated with them.
Although our main interest is atomic hydrogen clouds, and isolated H\,\textsc{i} clouds would in principle be ideal targets, regions consisting only of atomic hydrogen clouds generally lack independent distance constraints. 
We therefore focus on nearby molecular cloud complexes with known distances, but select their atomic envelopes and inter molecular cloud regions rather than the molecular interiors. 
Although molecular clouds are already evolved structures, their surroundings provide important laboratories for studying the evolution of H\,\textsc{i} gas and the formation of CNM.

The maximum recoverable angular scale of the SKA1-Mid AA$^{*}$ and AA4 is $\sim$0.5$^\circ$ at 1.4 GHz, corresponding to $\sim$1.3 pc at a distance of 150 pc. 
Using Equation (1), the Field length of the WNM is estimated to be $\sim$1.6 pc for $T = 10^4$ K and $n = 0.1$ cm$^{-3}$. This indicates that WNM structures will be significantly filtered out in observations with the SKA1-Mid AA$^{*}$ and AA4. 
By contrast, the maximum diameter of the CNM clumps identified in the GALFA-H\,\textsc{i} data, which have a spatial resolution of $\sim$0.17 pc, is $\sim$1.3 pc \citep{Matsuzuki2026}. 
This scale is comparable to the maximum recoverable angular scale of the SKA1-Mid AA$^{*}$ and AA4 at 1.4 GHz. 
Therefore, SKA-Mid AA$^{*}$ and AA4 observations will be sensitive to the full range of CNM structures in targets at 150 pc, from the initial $\lesssim$1000 AU seeds of CNM formation to evolved structures of order $\sim$1 pc that may represent the immediate precursors of molecular cloud formation. 
Such observations will provide critical insight into the formation of the CNM and its connection to the subsequent onset of molecular cloud formation.

\section{Summary}

The formation of the CNM is governed by thermal instability, turbulence, and magnetic fields, but their relative roles at the CNM formation scale remain uncertain.
Thermal instability provides the basic route by which diffuse atomic gas condenses into a cold phase, while the Field length defines the characteristic scale below which its growth is suppressed by thermal conduction.
Because the CNM is the immediate precursor of molecular clouds, understanding structure formation near the Field length is essential for clarifying the origin of the initial conditions for molecule formation.
This requires spatially resolved H\,\textsc{i} observations on scales comparable to and below the Field length.
Existing absorption-line measurements sample only isolated sightlines and cannot recover the morphology or internal kinematics of CNM structures.
High-resolution H\,\textsc{i} emission mapping with SKA1-Mid will instead make it possible to isolate CNM structures in both position and velocity, measure their basic properties statistically, and determine whether their velocity field is dominated by turbulence or by thermal instability and condensation-driven motions.
Combined with spatially extended H\,\textsc{i} Zeeman measurements, this approach will provide a new observational basis for establishing how thermal instability, turbulence, and magnetic fields shape the CNM and set the initial conditions for molecular cloud formation.

\section{Acknowlegdement}
This publication utilizes data from Galactic ALFA H\,\textsc{i} (GALFA H\,\textsc{i}) survey data set obtained with the Arecibo L-band Feed Array (ALFA) on the Arecibo 305m telescope. 
The Arecibo Observatory is operated by SRI International under a cooperative agreement with the National Science Foundation (AST-1100968), and in alliance with Ana G. Méndez-Universidad Metropolitana, and the Universities Space Research Association. 
The GALFA H\,\textsc{i} surveys have been funded by the NSF through grants to Columbia University, the University of Wisconsin, and the University of California. 

\bibliographystyle{abbrvnat-maxbibnames4}
\bibliography{chapter} % if your bibtex file is called example.bib

@article{Heiles2003a,
  title={The Millennium Arecibo 21 Centimeter Absorption-Line Survey. I. Techniques and Gaussian Fits},
  author={{Heiles}, C. and {Troland}, T. H.},
  year={2003},
  journal={\apjs},
  volume={145},
  pages={329--354},
}

@article{Heiles2003b,
  title={The Millennium Arecibo 21 Centimeter Absorption-Line Survey. II. Properties of the Warm and Cold Neutral Media},
  author={{Heiles}, C. and {Troland}, T. H.},
  year={2003},
  journal={\apj},
  volume={586},
  pages={1067--1093},
}

@article{Stanimirovic2014,
  title={Cold and Warm Atomic Gas around the Perseus Molecular Cloud. I. Basic Properties},
  author={{Stanimirovi$\acute{c}$}, S. and {Murray}, C. E. and {Lee}, M.-Y. and {Heiles}, C. and {Miller}, J.},
  year={2014},
  journal={\apj},
  volume={793},
  pages={id.132(16pp.)},
}

@article{Murray2015,
  title={The 21-SPONGE HI Absorption Survey I: Techniques and Initial Results},
  author={{Murray}, C. E. and {Stanimirovi$\acute{c}$}, S. and {Goss}, W. M. and {Dickey}, J. M. and {Heiles}, C. and {Lindner}, R. R. and {Babler}, B. and {Pingel}, N. M. and {Lawrence}, A. and {Jencson}, J. and {Hennebelle}, P.},
  year={2015},
  journal={\apj},
  volume={804},
  pages={id.89(21pp)},
}

@article{Murray2018,
  title={The 21-SPONGE HI Absorption Line Survey. I. The Temperature of Galactic HI},
  author={{Murray}, C. E. and {Stanimirovi$\acute{c}$}, S. and {Goss}, W. M. and {Heiles}, C. and {Dickey}, J. M. and {Babler}, B. and {Kim}, C.-G.},
  year={2018},
  journal={\apjs},
  volume={238},
  pages={id.14(25pp)},
}

@article{Faison2001,
  title={The Structure of the Cold Neutral Interstellar Medium on 10--100 AU Scales},
  author={{Faison}, M. D. and {Goss}, W. M.},
  year={2001},
  journal={\aj},
  volume={121},
  pages={2706--2722},
}

@article{Stanimirovic2018,
  title={Atomic and Ionized Microstructures in the Diffuse Interstellar Medium},
  author={{Stanimirovi$\acute{c}$}, S. and {Zweibel}, E. G.},
  year={2018},
  journal={\araa},
  volume={56},
  pages={489--540},
}

@article{Inoue2012,
  title={Formation of Turbulent and Magnetized Molecular Clouds via Accretion Flows of HI Clouds},
  author={{Inoue}, T. and {Inutsuka}, S.},
  year={2012},
  journal={\apj},
  volume={759},
  pages={id.35(14pp)},
}

@article{Zeeman1897,
  title={On the Influence of Magnetism on the Nature of the Light Emitted by a Substance},
  author={{Zeeman}, P.},
  year={1897},
  journal={\apj},
  volume={5},
  pages={332},
}

@article{Crutcher2019,
  title={Review of Zeeman Effect Observations of Regions of Star Formation},
  author={{Crutcher}, M. R. and {Athol}, J. K.},
  year={2019},
  journal={Frontiers in Astronomy and Space Sciences},
  volume={6},
  pages={66(18pp)},
}

@article{Yamamoto2003,
  title={High-Latitude Molecular Clouds in an H I Filament toward the MBM 53, 54, and 55 Complex: Existence of an H2 Cloud with Low CO Intensity},
  author={{Yamamoto}, H. and {Onishi}, T. and {Mizuno}, A. and {Fukui}, Y.},
  year={2003},
  journal={\apj},
  volume={592},
  pages={217--232},
}

@article{Kalberla2018,
  title={Properties of cold and warm H I gas phases derived from a Gaussian decomposition of HI4PI data},
  author={{Kalberla}, P. M. W. and {Haud}, U.},
  year={2018},
  journal={\aap},
  volume={619},
  pages={id.A58(20pp)},
}

@article{BenBekhti2016,
  title={HI4PI: A full-sky H I survey based on EBHIS and GASS},
  author={{HI4PI Collaboration:}, Bem Bekhti, N. and {Fl$\ddot{o}$}, L. and {Keller}, R. and {Kerp}, J. and {Lenz}, D. and {Winkel}, B. and {Bailin}, J. and {Calabretta}, M. R. and {Dedes}, L. and {Ford}, H. A. and {Gibson}, B. K. and {Haud}, U. and {Janowiecki}, S. and {Kalberla}, P. M. W. and {Lockman}, F. J. and {McClure-Griffiths}, N. M. and {Murphy}, T. and {Nakanishi}, H. and {Pisano}, D. J.},
  year={2016},
  journal={\aap},
  volume={594},
  pages={id.A116(15pp)},
}

@article{Kobayashi2020,
  title={Bimodal Behavior and Convergence Requirement in Macroscopic Properties of the Multiphase Interstellar Medium Formed by Atomic Converging Flows},
  author={{Kobayashi}, I. N. M. and {Inoue}, T. and {Inutsuka}, S. and {Tomida}, K. and {Iwasaki}, K. and {Tanaka}, E. I. K.},
  year={2020},
  journal={\apj},
  volume={905},
  pages={id.95(17pp)},
}

@article{Parker1953,
  title={Instability of Thermal Fields},
  author={{Parker}, E. N.},
  year={1953},
  journal={\apj},
  volume={117},
  pages={431--436},
}

@article{Wolfire2003,
  title={Neutral Atomic Phases of the Interstellar Medium in the Galaxy},
  author={{Wolfire}, G. M. and {McKee}, F. C. and {Hollenbach}, D. and {Tielens}, M. G. G. A.},
  year={2003},
  journal={\apj},
  volume={587},
  pages={278--311},
}

@article{Kortgen2015,
  title={Impact of magnetic fields on molecular cloud formation and evolution},
  author={{K$\ddot{o}$rtgen}, B. and {Banerjee}, R.}, 
  year={2015},
  journal={\mnras},
  volume={451},
  pages={3340--3353},
}

@article{McClure-Griffiths2023,
  title={Atomic Hydrogen in the Milky Way: A Stepping Stone in the Evolution of Galaxies},
  author={{McClure-Griffiths}, M. N. and {Stanimirovi$\acute{c}$}, S. and {Rybarczyk}, R. D.},
  year={2023},
  journal={\araa},
  volume={61},
  pages={19--63},
}

@article{Heiles2005,
  title={Magnetic Fields in Diffuse HI and Molecular Clouds},
  author={{Heiles}, C. and {Crutcher}, R.},
  year={2005},
  journal={Lecture Notes in Physics},
  volume={664},
  pages={137--182},
}

@article{PlanckCollaboration2018,
  title={Planck 2018 results XII. Galactic astrophysics using polarized dust emission},
  author={{Planck Collaboration}},
  year={2018},
  journal={\aap},
  volume={641},
  pages={A12(43pp)},
}

@article{Jarvis2015,
  title={Cosmology with SKA Radio Continuum Surveys},
  author={{Jarvis}, J. M. and {Bacon}, D. and {Blake}, C. and {Brown}, M. L. and {Lindsay}, S. and {Raccanelli}. A. and {Santos}, M. and {Schwarz}, D. J.},
  year={2015},
  journal={AASKA14},
  volume={215},
  pages={018(15pp)},
}

@article{Hale2025,
  title={MIGHTEE: the continuum survey Data Release 1},
  author={{Hale}, L. C. and {Heywood}, I. and {Jarvis}, J. M. and {Whittam}, H. I. and {Best}, N. P. and {An} F. and {Bowler}, A. A. R. and {Harrison}, I. and {Matthews}, A. and {Smith}, B. J. D. and {Taylor}, A. R. and {Vaccari}, M.},
  year={2025},
  journal={\mnras},
  volume={536},
  pages={2187--2211},
}

@article{Hennebelle2007,
  title={On the structure of the turbulent interstellar atomic hydrogen
I. Physical characteristics. Influence and nature of turbulence in a thermally bistable flow},
  author={{Hennebelle}, P. and {Audit}, E.},
  year={2007},
  journal={\aap},
  volume={465},
  pages={431--443},
}

@article{Saury2014,
  title={The structure of the thermally bistable and turbulent atomic gas in the local interstellar medium},
  author={{Saury}, E. and {Miville-Desch$\hat{e}$nes}, -A. M. and {Hennebelle}, P. and {Audit}, E. and {Schmidt}, W.},
  year={2014},
  journal={\aap},
  volume={567},
  pages={A16(22pp)},
}

@article{Heitsch2009,
  title={EFFECTS OF MAGNETIC FIELD STRENGTH AND ORIENTATION ON MOLECULAR CLOUD FORMATION},
  author={{Heitsch}, F. and {Stone}, M. J. and {Hartmann}, W. L.},
  year={2009},
  journal={\apj},
  volume={695},
  pages={248--258},
}

@article{Hennebelle2019,
  title={The Role of Magnetic Field in Molecular Cloud Formation and Evolution},
  author={{Hennebelle}, P. and {Inutsuka}, S.},
  year={2019},
  journal={FrASS},
  volume={6},
  pages={5(1--30)},
}

@article{Heiles2004,
  title={THE MILLENNIUM ARECIBO 21 CENTIMETER ABSORPTION-LINE SURVEY. III. TECHNIQUES FOR SPECTRAL POLARIZATION AND RESULTS FOR STOKES V},
  author={{Heiles}, C. and {Troland}, H. T.},
  year={2004},
  journal={\apjs},
  volume={151},
  pages={271--297},
}

@article{PlanckCollaboration2016,
  title={Planck intermediate results XLII. Large-scale Galactic magnetic fields},
  author={{Planck Collaboration}},
  year={2016},
  journal={\aap},
  volume={596},
  pages={A103(28pp)},
}

@article{Matsuzuki2026,
  title={Mapping of the Cold Neutral Medium via HI Phase Separation in an Atomic Cloud Undergoing Molecular Cloud Formation},
  author={{Matsuzuki}, Y. and {Yamamoto}, H. and {Tachihara}, K.},
  year={2026},
  journal={arXiv},
  volume={2603.14967},
  pages={19pp},
}

@article{Koyama2002,
  title={An Origin of Supersonic Motions in Interstellar Clouds},
  author={{Koyama}, H. and {Inutsuka}, S.},
  year={2002},
  journal={\apj},
  volume={564},
  pages={L97--L100},
}

@article{Seifried2020,
  title={From parallel to perpendicular - On the orientation of magnetic fields in molecular clouds},
  author={{Seifried}, D. and {Walch}, S. and {Weis}, M. and {Reissl}, S. and {Soler}, J. D. and {Klessen}, R. S. and {Joshi}, P. R.},
  year={2020},
  journal={\mnras},
  volume={497},
  pages={4196--4212},
}

@article{Weis2024,
  title={Properties of molecular clumps and cores in colliding magnetized flows},
  author={{Weis}, M. and {Walch}, S. and {Seifried}, D. and {Ganguly}, S.},
  year={2024},
  journal={\mnras},
  volume={532},
  pages={1262--1295},
}

@article{GrandaMunoz2025,
  title={The role of shocks and the velocity gradient in the relative orientation of the magnetic field and dense gas clouds},
  author={{Granda-Mu\~{n}oz}, G. and {V\'{a}zquez-Semadeni}, E. and {G\'{o}mez}, G. C.},
  year={2025},
  journal={\aap},
  volume={694},
  pages={A296(11pp)},
}

@article{Inoue2009,
  title={Two-Fluid Magnetohydrodynamics Simulations of Converging HI Flows in the Interstellar Medium. II. Are Molecular Clouds Generated Directly from a Warm Neutral Medium?},
  author={{Inoue}, T. and {Inutsuka}, S.},
  year={2009},
  journal={\apj},
  volume={704},
  pages={161--169},
}

@article{Inoue2008,
  title={Two-Fluid Magnetohydrodynamic Simulations of Converging HI Flows in the Interstellar Medium. I. Methodology and Basic Results},
  author={{Inoue}, T. and {Inutsuka}, S.},
  year={2008},
  journal={\apj},
  volume={687},
  pages={303--310},
}

@article{Audit2005,
  title={Thermal condensation in a turbulent atomic hydrogen flow},
  author={{Audit}, E. and {Hennebelle}, P.},
  year={2005},
  journal={\aap},
  volume={433},
  pages={1--13},
}

@article{Larson1981,
  title={Turbulence and star formation in molecular clouds.},
  author={{Larson}, R. B.},
  year={1981},
  journal={\mnras},
  volume={194},
  pages={809--826},
}

@article{Field1965,
  title={Thermal Instability.},
  author={{Field}, G. B.},
  year={1965},
  journal={\apj},
  volume={142},
  pages={531--567},
}

@article{Booth2026,
  title={A Three-dimensional Model for the Reversal in the Local Large-scale Interstellar Magnetic Field},
  author={{Booth}, A. R. and {Ordog}, A. and {Brown}, J-A. and {Landecker}, L. T. and {Hill}, S. A. and {West} L. J. and {Lei}, M. and {Clark}, E. S. and {Bacco}, A. and {Dickey}, M. J.},
  year={2026},
  journal={\apj},
  volume={997},
  pages={304(17pp)},
}

@article{Heiles1998,
  title={L204: A Gravitationally Confined Dark Cloud in a Strong Magnetic Environment},
  author={{Heiles}, C.},
  year={1988},
  journal={\apj},
  volume={324},
  pages={321--330},
}

@article{Heiles1999,
  title={Magnetic Fields, Pressures, and Thermally Unstable Gas in Prominent Hi Shells},
  author={{Heiles}, C.},
  year={1989},
  journal={\apj},
  volume={336},
  pages={808--821},
}

@article{Krumholz2009a,
  title={THE STAR FORMATION LAW IN ATOMIC AND MOLECULAR GAS},
  author={{Krumholz}, R. M. and {McKee}, F. C. and {Tumlinson}, J.},
  year={2009},
  journal={\apj},
  volume={699},
  pages={850--856},
}

@article{Krumholz2009b,
  title={THE ATOMIC-TO-MOLECULAR TRANSITION IN GALAXIES. II: HI AND H2 COLUMN DENSITIES},
  author={{Krumholz}, R. M. and {McKee}, F. C. and {Tumlinson}, J.},
  year={2009},
  journal={\apj},
  volume={693},
  pages={216--235},
}

@article{Kennicutt2012,
  title={Star Formation in the Milky Way and Nearby Galaxies},
  author={{Kennicutt Jr.} C. R. and {Evans II}, J. N.},
  year={2012},
  journal={\araa},
  volume={50},
  pages={531--608},
}

@article{Pattle2023,
  title={Magnetic Fields in Star Formation: from Clouds to Cores}, 
  author={{Pattle}, K. and {Fissel}, L. and {Tahani}, M. and {Liu}, T. and {Ntormousi}, E.},
  year={2023},
  journal={ASPC},
  volume={534},
  pages={193--232},
}

@article{Ching2022,
  title={An early transition to magnetic supercriticality in star formation}, 
  author={{Ching}, -C. T. and {Li}, D. and {Heiles}, C. and {Li}, -Y. Z. and {Qian}, L. and {Yue}, L. Y. and {Tang}, J. and {Jiao}, H. S.},
  year={2022},
  journal={\nat},
  volume={601},
  pages={49--52},
}

@article{Zucker2022,
  title={Star formation near the Sun is driven by expansion of the Local Bubble}, 
  author={{Zucker}, C. and {Goodman}, A. A. and {Alves}, J. and {Bialy}, S. and {Foley}, M. and {Speagle}, S. J. and {Gro$\beta$schedl}, J. and {Finkbeiner}, P. D. and {Burkert}, A. and {Khimey}, D. and {Swiggum}, C.},
  year={2022},
  journal={\nat},
  volume={601},
  pages={334--348},
}

@article{Yamamoto2006,
  title={Large-Scale CO Observations of a Far-Infrared Loop in Pegasus: Detection of a Large Number of Very Small Molecular Clouds Possibly Formed via Shocks}, 
  author={{Yamamoto}, H. and {Kawamura}, A. and {Tachihara}, K. and {Mizuno}, N. and {Onihsi}, T. and {Fukui}, Y.},
  year={2006},
  journal={\apj},
  volume={642},
  pages={307--318},
}

@article{Bracco2020,
  title={The multiphase and magnetized neutral hydrogen seen by LOFAR}, 
  author={{Bracco}, A. and {Jeli$\acute{c}$}, V. and {Marchal}, A. and {Turi$\acute{c}$}, L. and {Erceg}, A. and {Miville-Desch$\hat{e}$nes}, M. -A. and {Bellomi}, E.},
  year={2020},
  journal={\aap},
  volume={644},
  pages={id.L3(9pp)},
}

@article{Berat2026,
  title={Contribution of neutral gas to Faraday tomographic data at low frequencies}, 
  author={{Berat}, J. and {Miville-Desch$\hat{e}$nes}, M. -A. and {Bracco}, A. and {Hennebelle}, P. and {Scholtys}, J.},
  year={2026},
  journal={\aap},
  volume={708},
  pages={A245(23pp)},
}

@article{Boulanger2024,
  title={Associating LOFAR Galactic Faraday structures with the warm neutral medium}, 
  author={{Boulanger}, F. and {Gry}, C. and {Jenkins}, B. E. and {Bracco}, A. and {Erceg}, A. and {Jeli$\acute{c}$}, V. and {Turi$\acute{c}$}, L.},
  year={2024},
  journal={\aap},
  volume={687},
  pages={A102{12pp}},
}

@article{Robert2019,
  title={Anticipated Performance of the Square Kilometre Array -- Phase 1 (SKA1)},
  author={{Robert}, B. and {Anna}, B. and {Tyler}, B. and {Evan}, K. and {Jeff}, W.},
  year={2019},
  journal={arXiv},
  volume={1912.12699},
  pages={27pp},
}

\end{document}